\documentclass[aps,prl,amsmath,amssymb,superscriptaddress,showpacs,balancelastpage,10pt,reprint]{revtex4-1}

\usepackage[dvips]{graphicx}
\usepackage[colorlinks=true]{hyperref}
\usepackage[normalem]{ulem}

\graphicspath{{}}
\hypersetup{pdfpagemode = UseNone}

\newcommand{\setup}[1]{{\textsf{#1}}}

\newcommand{\Tc}{T_\mathrm{c}}

\newcommand{\nb}{n_\mathrm{b}}

\newcommand{\Pf}{P^\rho}
\newcommand{\Pb}{P^E}
\newcommand{\Pfb}{P}
\newcommand{\Tr}{\mathrm{Tr}}
\newcommand{\Ns}{S}
\DeclareRobustCommand\openone{\leavevmode\hbox{\small1\normalsize\kern-.33em1}}
\newcommand{\unity}{\openone}

\begin{document}
\title{Past quantum state analysis of the photon number evolution in a cavity}

\author{T.~Rybarczyk}
\author{S.~Gerlich}
\author{B.~Peaudecerf}
\author{M.~Penasa}
\affiliation{Laboratoire Kastler-Brossel, ENS, UPMC-Paris 6, CNRS, Coll\`ege de France, 11 place Marcelin Berthelot, 75005 Paris, France.}
\author{B.~Julsgaard}
\author{K.~M\o{}lmer}
\affiliation{Department of Physics and Astronomy, Aarhus University, Ny Munkegade 120, DK-8000 Aarhus C, Denmark.}
\author{S.~Gleyzes}
\author{M.~Brune}
\author{J.M.~Raimond}
\author{S.~Haroche}
\author{I.~Dotsenko}
\affiliation{Laboratoire Kastler-Brossel, ENS, UPMC-Paris 6, CNRS, Coll\`ege de France, 11 place Marcelin Berthelot, 75005 Paris, France.}
\email{igor.dotsenko@lkb.ens.fr}
\date{\today}

\begin{abstract}
A quantum system can be monitored through repeated interactions with meters, followed by their detection. The state of the system at time $t$ is thus conditioned on the information obtained until that time. More insight in the state dynamics is provided, however, by the past quantum state (PQS) [S. Gammelmark {\sl et al.} Phys.~Rev.~Lett. {\bf 111}, 160401 (2013)]. It relies on all aspects of the system evolution which are recorded in the past {\sl and} in the future of $t$. Using PQS analysis for the quantum non-demolition photon number counting in a cavity, we can reveal information hidden in the standard approach and resolve a wider range of number states. This experiment demonstrates the strong potential of PQS analysis.
\end{abstract}

\maketitle

Journalists' comments on the present time $t$ are based on their knowledge of the present and of the past. It is sometimes difficult for them to single out the relevant events from the random noise of daily news, or to lift ambiguities between equally probable interpretations. The situation of historians, working at a future time~$T$ is quite different. They base their insights onto all events up to their own time. Their knowledge of the future of $t$ is instrumental in sorting out the relevant events and in lifting ambiguities.

Similar considerations apply to the monitoring of a quantum system, probed by meters performing repeated generalized measurements \cite{BraginskyBook}. The common approach to state estimation at time $t$ is to follow the evolution of the density matrix of the system $\rho$ from $0$ to $t$, due to the intrinsic evolution (including relaxation) and to the interaction with the meters. This is the journalist's perspective. It is prone to noise induced by quantum statistical fluctuations of the meters output. Ambiguities may also arise, particularly when the meter is an interferometric device, a number of system states leading then to the same meter reading.

State estimation at time $t$ is considerably improved if we adopt the historian's perspective, and supplement our knowledge with meter readings from $t$ to $T$. Noise due to statistical fluctuations can be significantly reduced. Ambiguities can be lifted when the candidate state assignment at $t$ based on the ordinary analysis is in blatant contradiction with future evolution.

A recent paper \cite{Gammelmark13} provides a simple formalism describing the best estimate about the system at $t$ from information gathered in the past {\sl and} in the future of $t$. The meter readings  and the intrinsic system's evolution between $0$ and $t$ provide, in a {\sl forward} analysis, the density operator at $t$, $\rho(t)$. We can then compute the ordinary probability, $P^\rho(n,t)$, for observing at $t$ the result $n$ in any quantum measurement described by a set of positive-operator valued measures (POVM) $\{\hat\Omega^\dag_n\hat\Omega_n\}$:
   \begin{equation}\label{eq:standard}
		P^\rho(n,t) = \Tr\big[\hat\Omega^\dag_n\hat\Omega_n\rho(t)\big].
    \end{equation}
 
A better estimate of the measurement outcome at $t$, taking also into account the meters recorded from $t$ to $T$, is described by the \textit{past quantum state} (PQS), a pair of operators $\{\rho(t),E(t)\}$. The ``effect matrix", $E(t)$, includes data gathered between $t$ and $T$ in a time-reversed (``backwards") sequence, as well as the time-reversed intrinsic evolution. The probability of a measurement outcome $n$ at time $t$ can be computed from the PQS as:
    \begin{equation}\label{eq:Klaus}
		P(n,t) = \frac{\Tr\big[\hat\Omega_n\rho(t)\hat\Omega^\dag_n E(t)\big]}{\sum_m\Tr\big[\hat\Omega_m\rho(t)\hat\Omega^\dag_m E(t)\big]}.
    \end{equation}

The PQS formalism has already been used for the continuous monitoring of a two-level atom in a cavity \cite{Gammelmark14}, leading to a considerable improvement on the system's parameter estimation. It has also been applied to weak measurements performed at $t$ on a two-level system undergoing projective measurements at $0$ and $T$ \cite{Huard14}, in close connection with Aharonov's ``weak values" \cite{Aharonov91,Wiseman02}.

We present here an experiment applying the PQS approach to a system with a high-dimension Hilbert space. We study the photon number, $n$, in a cavity repeatedly probed by ``meter" atoms interacting with it in the dispersive regime~\cite{HarocheBook}. The atoms experience a photon-number-dependent light shift, which is read out using a Ramsey atomic interferometer. Hence, the meter reading is an ambiguous periodic function of $n$. This feature, exhibiting vividly the interest of PQS, was absent in~\cite{Gammelmark14,Huard14}. We show that the photon number estimation based on the PQS is much more reliable than the standard one and that ambiguities in the state assignment are lifted. From the PQS we extract information which is ordinarily hidden. This experiment demonstrates the wide potential of the PQS approach.

    \begin{figure}[t]
     \includegraphics[width=7cm]{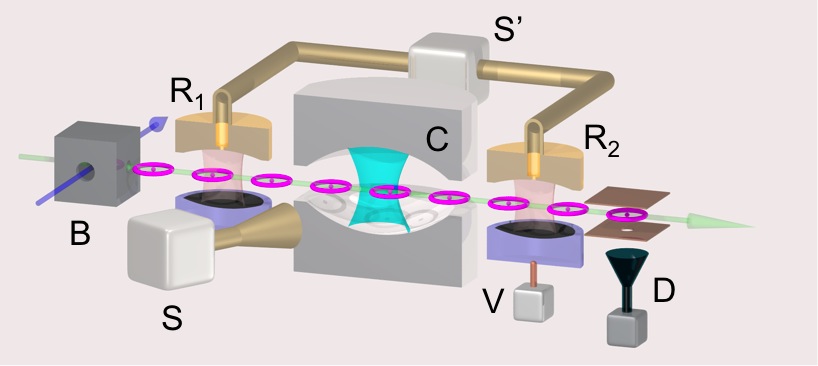}
     \caption{Scheme of the experimental setup. See the text for details.}
     \label{fig:setup}
    \end{figure}
    
The scheme of the set-up is presented in Fig.~\ref{fig:setup}. The microwave field is stored in a high-$Q$ superconducting cavity \setup{C} resonant at $\omega_c/2\pi=51$~GHz \cite{Kuhr07}. The cavity, cooled down to $0.8$~K, has an energy damping time $\Tc=65$~ms. It is repeatedly probed by circular Rydberg atoms, excited in \setup{B} from a Rubidium thermal beam. The atomic transition frequency $\omega_\mathrm{a}$ between the circular states with principal quantum numbers $50$ (state $\vert g\rangle$) and $51$ (state $\vert e\rangle$) is close to $\omega_\mathrm{c}$. Atomic samples cross the cavity mode every $T_\mathrm{a}=86\,\mu$s. The atomic state is finally measured in the field-ionization detector \setup{D}. On the average, we detect 0.28 atom per sample. 

The dispersive interaction of an atom with $n$ photons in \setup{C} changes the relative phase between $\vert g\rangle$ and $\vert e\rangle$ by $\varphi(n) \approx \varphi_0 (n+1/2)$, where  $\varphi_0$ is the phase shift per photon \cite{HarocheBook}. Information on $\varphi(n)$, and thus on $n$, is read out using a Ramsey interferometer, made up of two low-$Q$ cavities, \setup{R$_1$} and \setup{R$_2$}, sandwiching  \setup{C}. They induce $\pi/2$ classical Rabi pulses between $\vert g\rangle$ and $\vert e\rangle$. The conditional probability to detect the atom in state $a\in\{g,e\}$ (the ``Ramsey fringes" signal) is
    \begin{equation}\label{eq:fringe}
        P(a|\phi,n) =\left\{1+jA+jB\,\sin[\varphi(n)-\varphi_0/2-\phi] \right\}/2,
    \end{equation}
where $j=1$ ($-1$) for $a=g$ ($e$). The finite offset and the reduced contrast of the fringes ($A=0.03$ and $B=0.71$, respectively) are due to experimental imperfections. The Ramsey interferometer phase $\phi$ is controlled via a transient Stark shift of $\omega_a$ produced by the electric potential \setup{V} applied across \setup{R$_2$}. 

We set here the average phase shift per photon to be $\varphi_0\approx\pi/4$ by adjusting the atom-cavity detuning. In order to optimize photon number discrimination, we alternate the Ramsey interferometer phase $\phi$ between four values, approximately equal to $0$, $\pi/4$, $\pi/2$, and $3\pi/4$. Nevertheless, due to the periodicity of $P(a|\phi,n)$, the measurement is {\sl a priori} unable to distinguish $n$ photons from $n+8$~\cite{Guerlin07}.

The photon-number distribution $P(n,t)$ at time $t$ is obtained by replacing in Eq.~\eqref{eq:Klaus} the POVM operators $\Omega_n$ by the projectors $\vert n\rangle\langle n\vert$ on the Fock states:
    \begin{equation}\label{eq:forward_backward}
		\Pfb(n,t) = \frac{\Pf(n,t)\Pb(n,t)}{\sum_m \Pf(m,t)\Pb(m,t)},
    \end{equation}
where $\Pf(n,t)=\rho_{nn}(t)$ and $\Pb(n,t)=E_{nn}(t)$ are the diagonal elements of $\rho$ and $E$ in the $\{|n\rangle\}$ basis. The PQS distribution, $\Pfb(n,t)$, which includes all available information, is the normalized product of the forward, $\Pf(n,t)$, and backward, $\Pb(n,t)$, distributions. The forward one represents knowledge on $n$ based on all information acquired before $t$. The backward distribution reflects information provided by meters recorded between $t$ and $T$.

The forward distribution, $\Pf(n,t)$, reads at time $t_s=s T_\mathrm{a}$ immediately after the detection of the $s$th sample \cite{Peaudecerf13}
    \begin{equation}\label{eq:forward}
        \Pf(n,t_s) = \mathbb{M}_s\mathbb{T}\ \mathbb{M}_{s-1}\mathbb{T}\ \ldots \ \mathbb{M}_{1}\mathbb{T}\  \Pf(n,0)/{\cal N}^\rho_s
    \end{equation}
where ${\cal N}^\rho_s$ is a normalization and $\Pf(n,0)$ is assumed to be a uniform distribution for an initially unknown field, $\Pf(n,0) = 1/N$ ($N$ is the  Hilbert space size, chosen to be large enough). 

The operators $\mathbb{M}_i$ describe the update of the photon number distribution due to the detection of the $i$-th sample. Within a normalization, this update is deduced from Bayes' law and the action of  $\mathbb{M}_i$ on a probability distribution $p(n)$ is
 \begin{equation}\label{eq:Bayes}
       p(n)\longrightarrow \mathbb{M}_i p(n)= P(a_i|\phi_i,n) p(n).
    \end{equation}    
If no atom has been detected in the $i$-th sample, we must replace the linear operator $\mathbb{M}_i$ by the identity $\unity$. 
 
In the time interval $T_\mathrm{a}$ between the detection of samples $i-1$ and $i$, the photon distribution is also updated by the effect of cavity relaxation. Here, $T_\mathrm{a}\ll \Tc$, and this update can be written in terms of the linear operator $\mathbb{T}$ acting on a distribution $p(n)$ as
    \begin{equation}\label{eq:relaxation}
        p(n)\longrightarrow \mathbb{T} p(n) = \sum_{m}(I + T_\mathrm{a} K_{n,m}) p(m),
    \end{equation}
where $K_{n,n}\!=\!-\kappa[(1\!+\!\nb)n\!+\!\nb(n\!+\!1)]$, $K_{n,n+1}=\kappa (1\!+\!\nb)(n\!+\!1)$, $K_{n,n-1}=\kappa \nb n$, all the other coefficients being $0$ \cite{Brune08}. In these expressions, $\kappa=1/\Tc$ is the field energy damping rate and $\nb=0.074$ is the thermal photon number.

    \begin{figure*}[ht]
     \includegraphics[width=0.8\textwidth]{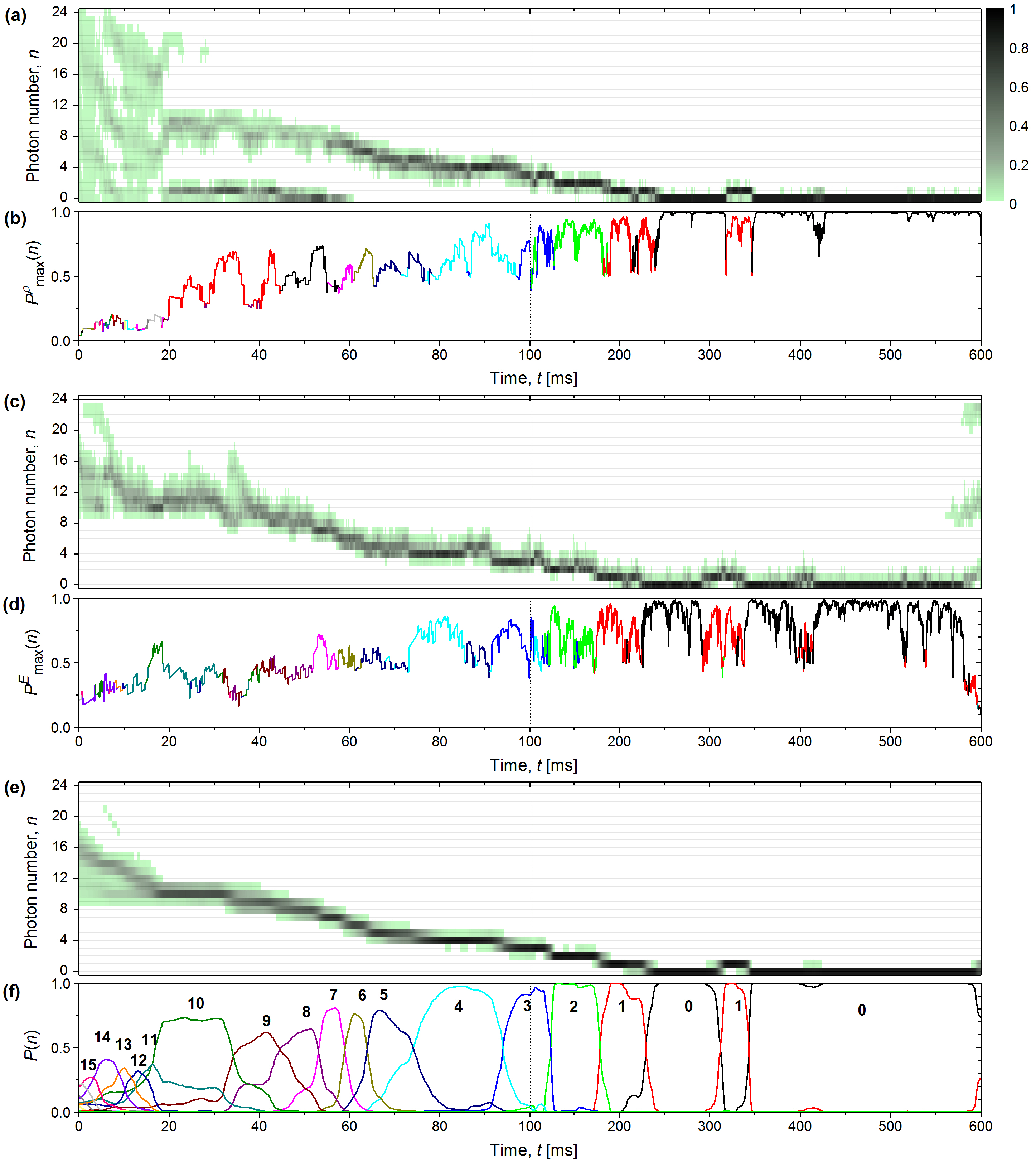}
     \caption{(Color online) Evolution of the estimated photon-number distributions. Panels (a)-(b), (c)-(d), and (e)-(f) show the forward, backward and PQS distributions respectively in a typical realization of the experiment. Panels (a), (c) and (e) present data with the color shade scale given in (a). Panels 
 (b), (d), and (f) give the explicit evolution of the photon number probabilities.  For the sake of clarity, we plot only the probability of the most likely photon number in panels (b) and (d). This number is given by the lines color code, defined by the labels in panel (f). Note that the time scale on the horizontal axis is changed by a factor of 5 at $t=100$~ms for all panels (vertical dotted line).}
     \label{fig:FBA}
    \end{figure*}
   
The equation for the effect matrix $E(t)$, evolving backwards from time $T$, can be similarly reduced to an (adjoint) update equation for its diagonal elements:
\begin{equation}\label{eq:backward}
        \Pb(n,t_s) = \mathbb{T}^\dagger\mathbb{M}_{s+1}\ \mathbb{T}^\dagger\mathbb{M}_{s+2}\ \ldots \ \mathbb{T}^\dagger\mathbb{M}_{S}\  \Pb(n,T)/{\cal N}^E_s
    \end{equation}
Here, $T=ST_a$ is the detection time of the final sample $S$ and $\Pb(n,T)=1/N$. This distribution is evolved backwards in time and takes into account all detection results from $\Ns$ back to $s+1$. The relaxation operator used in the forward analysis is replaced by its adjoint, $\mathbb{T}^\dag$,  to describe the effect of time-reversed cavity damping between detection events. 
  
We have performed two experiments illustrating the interest of the PQS approach. In the first one, we inject photons in the cavity and then send a sequence of $\Ns=7\,000$ meter samples (total duration $T=602$~ms). For the sake of experimental convenience we prepare initially a 12-photon coherent state. However, this information is discarded in our analysis in order to compare different photon number estimation approaches independently of any a priory information on the field preparation. We thus set $\Pf(n,0)=1/N$ with $N=25$.

Figure~\ref{fig:FBA} shows a single realization of the experiment. We plot the forward, backward, and PQS photon number distributions versus time. The ``noise" observed in $\Pf$ and $\Pb$ is mainly due to the statistical fluctuations of the random atomic detections (represented by the $\mathbb{M}_i$ operators, which can cause considerable changes in the photon number distribution). Between actual meter detections (occurring each 0.3~ms on the average), the estimated distributions evolve smoothly under cavity relaxation ($\mathbb{T}$ and $\mathbb{T}^\dag$ operators are close to $\unity$). 

During the first 20 ms, the forward distribution in Fig.~\ref{fig:FBA}(a) exhibits three significant maxima, separated by the $n=8$ period of the meter interferometric read-out set by the choice $\varphi_0\simeq \pi/4$. Between 20 and 40~ms, only two maxima are left around 1 and 9 photons, relaxation making high photon numbers less and less likely. At $\simeq$40~ms, the most probable number jumps from one to zero. Suddenly, at 60~ms, it jumps from 0 up to 7, before relaxing towards zero in a series of downwards jumps. The large upwards jump at 60~ms is an extremely unlikely event. In fact, the cavity contained most probably $9$ photons around $t=40$~ms, a state identified by the forward analyzis as $1$. This qualitative example illustrates how detection results obtained after $t$ can radically change state estimation at $t$. 

The backward distribution $\Pb(n,t)$ in Fig.~\ref{fig:FBA}(c)-(d) reflects the periodicity of $P(a|\phi,n)$ only at the end of the experimental sequence. Starting from time $T$, it quickly converges towards a mixture of $n$'s close to 0 modulo 8. Then, at earlier times, the combination of time-reversed decoherence and meter readings makes $0$ the most probable photon number. From then on, continuing backwards in time, the photon number increases and never shows abrupt jumps by $\pm8$, as was the case for~$\Pf(n,t)$.

Figures~\ref{fig:FBA}(e)-(f) show $\Pfb(n,t)$, the normalized product of $\Pb(n,t)$ and $\Pf(n,t)$. The first striking observation is the impressive reduction of the noise [compare panel (f) to (b) and (d)]. In contrast to $\Pf(n,t)$ and $\Pb(n,t)$, $\Pfb(n,t)$ includes all measurement and relaxation operators. Two consecutive values, at times $t_s$ and $t_{s+1}$, only differ by the arrangement of these operators in the evaluation of Eq.~\ref{eq:forward_backward}.

Moreover, the population of the most probable photon number is closest to one for $\Pfb(n,t)$. The photon number is thus determined with a higher fidelity. The  times at which the quantum jumps occur are also defined in a much clearer way. Noise makes this definition much less reliable for  $\Pf(n,t)$ and $\Pb(n,t)$. 

     \begin{figure}[t]
     \includegraphics[width=7cm]{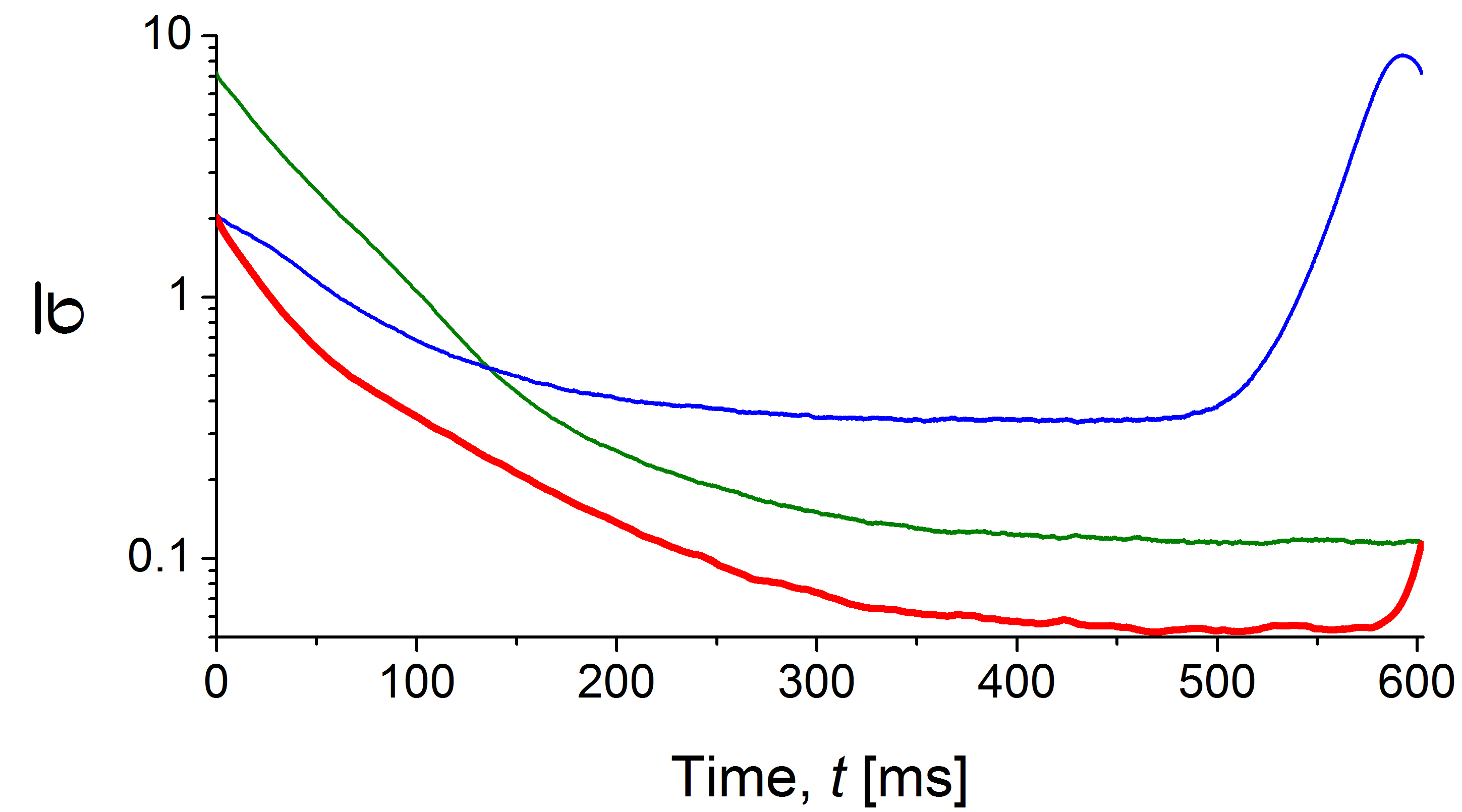}
     \caption{(Color online) Evolution of the average standard deviation $\overline \sigma $ of the forward (green line), backward (blue line), and PQS (thick red line) photon-number distributions.}
     \label{fig:std}
    \end{figure}

Finally, panels (e)-(f) clearly show that the ambiguity in the photon number is completely lifted revealing information hidden in the standard forward analysis. We can monitor the  series of quantum jumps from the initial photon number, around $15$ in this example, down to vacuum. 

We have analyzed $6\,000$ realizations of this experiment. The average standard deviations of the forward, backward and PQS distributions are shown in Fig.~\ref{fig:std}. The initial forward and final backward distributions being uniform, the corresponding deviations, $\overline\sigma[\Pf]$ and $\overline\sigma[\Pb]$, are the largest. For the same reason, $\overline\sigma[\Pfb]$ coincides with $\overline\sigma[\Pb]$ at $t=0$, and with $\overline\sigma[\Pf]$ at $t=T$. At all other times, $\overline \sigma[\Pfb] $ is the lowest. This clearly shows that the PQS provides a better photon-number estimation than the standard approach.

This first experiment does not give any indication about the precision of the determination of the  quantum jump times. In a second experiment, we use the PQS approach to detect a quantum jump induced on purpose at a well-defined time. The experimental sequence involves three parts. Starting with the residual thermal field, we first send $4\,000$ meter samples. We then induce a photon creation quantum jump by sending a single sample prepared in  $\vert e \rangle$. Using the Stark effect in an electric field pulse applied across the cavity mirrors, we tune this sample in resonance with \setup{C}, leading to atomic emission with a  high probability. We then resume the measurement of the field with $4\,000$ new meter samples. The experiment is repeated $16\,320$ times. We select the $2\,962$ realizations with exactly one atom detected in state $\vert g \rangle$ in the resonant sample. We thus isolate the sequences in which the quantum jump has most likely been successfully induced.

	\begin{figure}[t]
		\includegraphics[width=7cm]{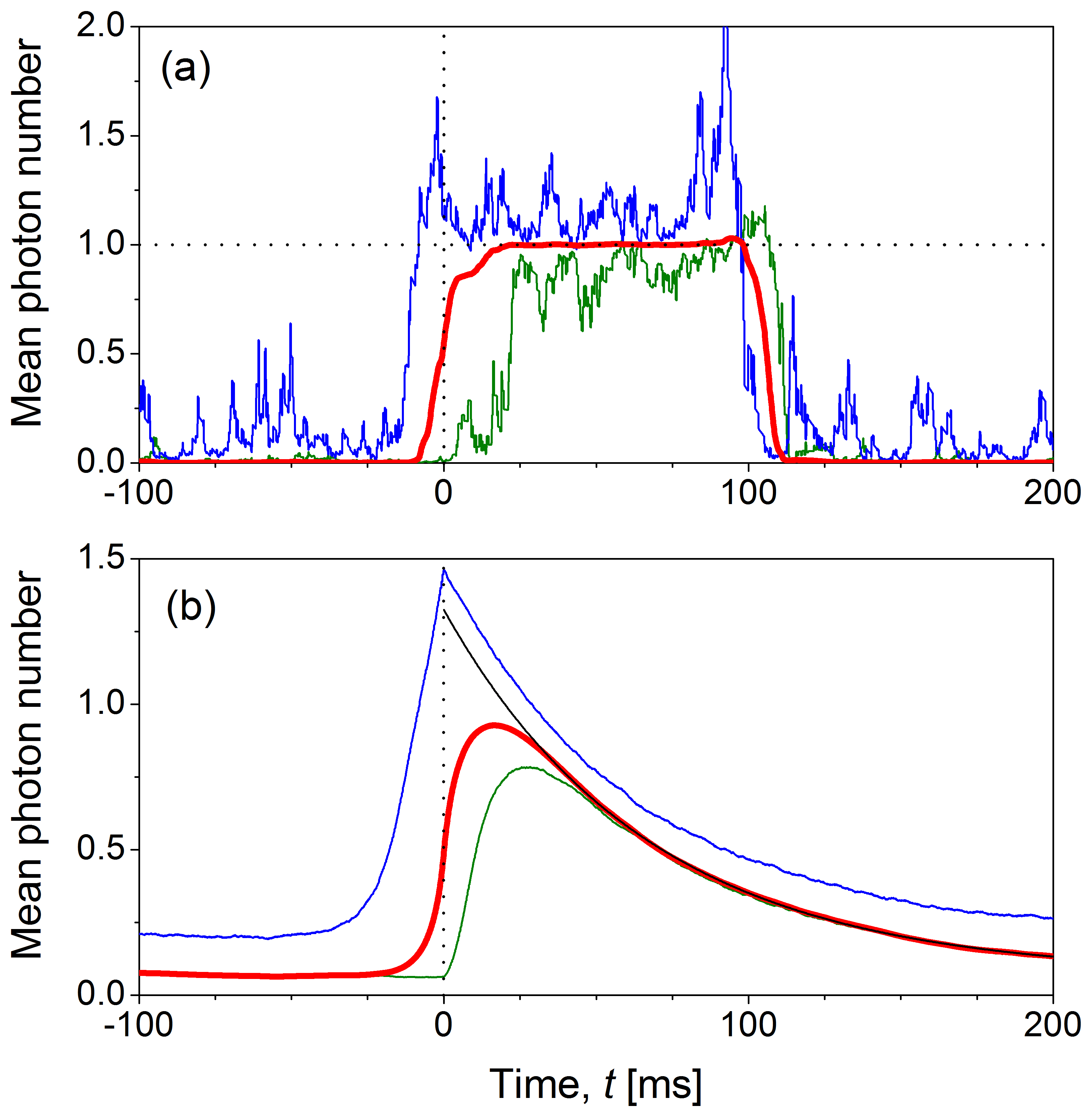}
		\caption{(Color online) Detection of a photon creation quantum jump induced at $t=0$. (a) Evolution of the mean photon number in the  forward (green), backward (blue), and PQS (thick red) analysis for a single realization. (b) Average over $2\,962$ realizations. The black line is an exponential fit to the PQS result.}
		\label{fig:QJump}
	\end{figure}
	
Figure \ref{fig:QJump} shows the mean photon number obtained from the forward (green), backward (blue), and PQS (red thick) analyses. Panel (a) presents a single realization. The time origin, $t=0$, corresponds to the induced jump. We only represent an interval of interest close to the jump. As expected, the forward (backward) measurement detects the induced jump later (earlier) than its real occurrence time. The PQS analysis gives a much better estimate: the jump time, defined as when the mean photon number crosses the $0.5$ level, is  much closer to $0$. The standard deviation of all jump times is $4.4$~ms, corresponding to $13$ detected atoms.

Figure \ref{fig:QJump}(b) shows an average over all selected realizations. The PQS (red) curve crosses the $0.5$ level at $0.1$~ms, a deviation from zero shorter than the delay between two detected atoms, demonstrating unbiased estimation of the jump time. Note that the average jump detection with standard analysis (green curve) is delayed by about 10~ms.

The photon is lost from the cavity after a time that varies in different realizations in accordance with an exponential decay. The black line in Fig.~\ref{fig:QJump}(b) is an exponential fit of the PQS data obtained after $35$~ms. The fit parameters are the decay constant, $67$~ms (close to the independently determined $\Tc$), an offset of $0.068$ photons (close to $n_\mathrm{b}$), and an amplitude of $1.27$ at $t=0$. This initial photon number is higher than $1$ due to the events in which two atoms in the resonant sample (one remaining undetected) inject two photons into \setup{C}. This value is in excellent agreement with a prediction based on the efficiency of \setup{D} (30\%) and the Poisson distribution of the atom number in each sample.

We have applied the past quantum state formalism to the determination of the photon number in a cavity. By using the results of all dispersive meter atoms, before and after time $t$, we get a much better estimate of the photon number and of its quantum jumps than with the standard approach, which uses only information available at and before $t$. By removing ambiguities in the photon number, we also access information hidden behind the periodicity of the interferometric meter read-out in the standard approach.

This experiment demonstrates the wide potential of PQS analysis. The method can be transposed in a variety of contexts, and it is highly relevant to quantum-enabled metrology, in which quantum state estimation is a key feature.

\begin{acknowledgments}
The authors acknowledge support from  European Research Council (DECLIC project), European Community (SIQS project) and Agence Nationale de la Recherche (QUSCO-INCA project).
\end{acknowledgments}

\bibliographystyle{plain}

\begin{thebibliography}{}

\end{thebibliography}


\begin{thebibliography}{99}

\bibitem{BraginskyBook}
V.~B.~Braginsky and F.~Y.~Khalili. {\em Quantum Measurement}, Cambridge University Press (1999)

\bibitem{Gammelmark13}
S. Gammelmark, B. Julsgaard, and K. M\o{}lmer, Phys.~Rev.~Lett. {\bf 111}, 160401 (2013).

\bibitem{Gammelmark14}
S.~Gammelmark, K. M\o{}lmer, W.~Alt, T.~Kampschulte, and D.~Meschede, Phys. Rev. A {\bf 89}, 043839 (2014).

\bibitem{Huard14}
P.~Campagne-Ibarcq, L.~Bretheau, E.~Flurin, A.~Auff\`eves, F.~Mallet, and B.~Huard, Phys. Rev. Lett. {\bf 112}, 180402 (2014).

\bibitem{Aharonov91}
Y. Aharonov and L. Vaidman, J. Phys. A {\bf 24}, 2315 (1991)

\bibitem{Wiseman02}
H. M. Wiseman, Phys. Rev. A {\bf 65}, 032111 (2002)

\bibitem{HarocheBook}
S.~Haroche and J.M.~Raimond. {\em Exploring the Quantum: atoms, cavities and photons}, Oxford University Press, Oxford (2006)

\bibitem{Kuhr07}
S.~Kuhr {\sl et al.}, Appl. Phys. Lett. {\bf 90}, 164101 (2007).

\bibitem{Guerlin07}
C.~Guerlin {\sl et al.}, Nature (London) {\bf 448}, 889 (2007).

\bibitem{Peaudecerf13}
B.~Peaudecerf {\sl et al.}, Phys. Rev. A {\bf 87}, 042320 (2013).

\bibitem{Brune08}
M. Brune {\sl et al.}, Phys. Rev. Lett. {\bf 101}, 240402 (2008).

\end{thebibliography}

\end{document}